\documentclass[fleqn,12pt]{article}

\usepackage{amsfonts}
\usepackage{graphics,graphicx}
\usepackage{amsmath}
\usepackage{amssymb}

\usepackage{color}

\usepackage{amsfonts,amssymb,cite}

\def\onecol{\onecolumn \mathindent 2em}

\def\noi{\noindent}

\newcommand{\Title}[1]{\noi {{\Large\bf #1}}\\[1ex]}

\newcommand{\Author}[2]{\noi{\bf #1}\\[2ex]\noi{\normalsize\it #2}\\}

\newcommand{\foom}[1]{\protect\footnotemark[#1]}

\def\email#1#2{\footnotetext[#1]{e-mail: #2}\addtocounter{footnote}{1}}

\def\nqq{\hspace*{-2em}}
\def\nhq{\hspace*{-0.5em}}

\def\cm{\hspace*{1cm}}
\def\inch{\hspace*{1in}}

\def\Jl#1#2{#1 {\bf #2},\ }

\def\ApJ#1 {\Jl{Astroph. J.}{#1}}
\def\CQG#1 {\Jl{Class. Quantum Grav.}{#1}}
\def\DAN#1 {\Jl{Dokl. AN SSSR}{#1}}
\def\GC#1 {\Jl{Grav. Cosmol.}{#1}}
\def\GRG#1 {\Jl{Gen. Rel. Grav.}{#1}}
\def\JETF#1 {\Jl{Zh. Eksp. Teor. Fiz.}{#1}}
\def\JETP#1 {\Jl{Sov. Phys. JETP}{#1}}
\def\JHEP#1 {\Jl{JHEP}{#1}}
\def\JMP#1 {\Jl{J. Math. Phys.}{#1}}
\def\NPB#1 {\Jl{Nucl. Phys. B}{#1}}
\def\NP#1 {\Jl{Nucl. Phys.}{#1}}
\def\PLA#1 {\Jl{Phys. Lett. A}{#1}}
\def\PLB#1 {\Jl{Phys. Lett. B}{#1}}
\def\PRD#1 {\Jl{Phys. Rev. D}{#1}}
\def\PRL#1 {\Jl{Phys. Rev. Lett.}{#1}}

\def\al{&\nhq}
\def\lal{&&\nqq {}}
\def\eq{Eq.\,}

\def\beq{\begin{equation}}
\def\eeq{\end{equation}}
\def\bear{\begin{eqnarray}}
\def\bearr{\begin{eqnarray} \lal}
\def\ear{\end{eqnarray}}
\def\earn{\nonumber \end{eqnarray}}

\def\nnn{\nonumber\\ \lal }

\def\yy{\\[5pt] {}}
\def\yyy{\\[5pt] \lal }
\def\eql{\al =\al}

\def\d{\partial}

\def\const{{\rm const}}
\def\eps{\varepsilon}

\def\sph{spherically symmetric}
\def\ssph{static, spherically symmetric}

\def\rf#1{\eqref{#1}}

\begin{document}
\onecol

\Title{Rastall's theory of gravity: Spherically symmetric solutions\yy
 and the stability problem}

\Author{K. A. Bronnikov\foom 1}
	{\small
	Center fo Gravitation and Fundamental Metrology, VNIIMS, 
		Ozyornaya ul. 46, Moscow 119361, Russia;\\
	 Institute of Gravitation and Cosmology, RUDN University, 
		ul. Miklukho-Maklaya 6, Moscow 117198, Russia;\\
	National Research Nuclear University ``MEPhI'', 
		Kashirskoe sh. 31, Moscow 115409, Russia}
   
 \Author{J\'ulio C. Fabris\foom 2}
 	{\small N\'ucleo Cosmo-ufes\ \&\ Departamento de F\'{\i}sica, CCE,
	Universidade Federal do Esp\'{\i}­rito Santo,\\
	Vit\'oria, ES, CEP29075-910, Brazil\\
	National Research Nuclear University ``MEPhI'', 
		Kashirskoe sh. 31, Moscow 115409, Russia}
 
 \Author{Oliver F. Piattella}
 	{\small N\'ucleo Cosmo-ufes\ \&\ Departamento de F\'{\i}­sica, CCE,
	Universidade Federal do Esp\'{\i}rito Santo,\\
	Vit\'oria, ES, CEP29075-910, Brazil}
 	
 \Author{Denis C. Rodrigues}
 	{\small PPGFis, CCE,
	Universidade Federal do Esp\'{\i}­rito Santo, Vit\'oria, ES, CEP29075-910, Brazil}	
 
 \Author{Edison C.O. Santos}
 {\small PPGCosmo, CCE,
	Universidade Federal do Esp\'{\i}­rito Santo, Vit\'oria, ES, CEP29075-910, Brazil}

\begin{abstract}
  We study the stability of static, spherically symmetric solutions of Rastall's theory
  in the presence of a scalar field with respect to spherically symmetric perturbations. 
  It is shown that the stability analysis is inconsistent in the sense that linear time-dependent 
  perturbations cannot exist, and we can conclude that these solutions are stable. Possible 
  reasons for this  inconsistency are discussed.
\end{abstract}

\email 1 {kb20@yandex.ru}
\email 2 {julio.fabris@cosmo-ufes.org}

\section{Introduction}

  One of the fundamental pillars of the General Relativity (GR) theory of gravity is the 
  divergence-free property of the Einstein tensor, leading to the usual conservation laws
  for the matter sources of gravity, expressed in the zero divergence of the
  energy-momentum tensor.

  Rastall's theory \cite{rastall} is a non-Lagrangian theory of gravity which gives up the above 
  property of GR and allows for an energy-momentum tensor of matter which has a nonzero
  divergence. The main original argument for such a radical departure from GR is that, 
  strictly speaking, the conservation laws have only be tested in flat space, and their 
  generalization to curved space-time may be more problematic than is usually supposed. 
  The departure from the usual GR conservation law is parametrized, in the Rastall theory 
  by the dimensionless parameter $\lambda$, such that when $\lambda = 1$, GR is recovered.

  The Rastall theory has been applied in many different contexts. In cosmology, some studies 
  have been performed for the early \cite{oliver} and also later universe \cite{ce}. In some of 
  these applications, a self-interacting scalar field has been considered. A curiosity about the 
  self-interacting scalar field in the Rastall theory is that the corresponding speed of sound
   $c_s^2$ can be zero for some value of $\lambda$ \cite{liddle}, unlike the GR situation
   where always $c_s^2 = 1$  (in unities of the velocity of light). Moreover, It is possible to 
   construct a cosmological model entirely similar to the standard $\Lambda$CDM model 
   for the background and linear perturbations, but different for nonlinear perturbations. 

  In Ref. \cite{kirillr} black-hole solutions using Rastall's theory in the presence of a scalar field
  have been obtained. Two classes of exotic black holes have been identified: one where the
  singularities are located at two spatial infinities separated by horizons; the other where there 
  are two horizons of infinite area connected by a wormhole, and the spatial infinity is essentially
  formed by a cosmological singularity. It is very interesting that these structures are geometrically
  identical to those found using k-essence theories, but with a different behavior of the scalar 
  fields \cite{kirillk}. In Ref. \cite{kirille} the equivalence (or duality) between these structures was 
  discussed in detail: we have formulated the assumptions under which the two fundamentally 
  different theories lead to space-times with the same metric.

  A question that emerges immediately concerns the stability of these exotic black hole-type
  structures. In Ref. \cite{kirills} the stability of the corresponding k-essence solutions 
  under spherically symmetric perturbations was analyzed, and their instabilities have been 
  proved. In the present paper, we undertake a similar analysis for the Rastall black hole-type
  solutions. A surprising aspect of this study is that the linear perturbation analysis turns out to 
  be inconsistent, and the stability or instability of the solutions cannot be determined. 
  More precisely, the linear perturbation equations cannot have time-dependent 
  solutions. Such a result looks similar to the cosmological perturbation analysis for a
  self-interacting scalar field, which has also revealed an inconsistency, unless 
  new degrees of freedom of matter are added to the previously existing matter 
  content of the Universe  \cite{daouda}. This may point out at an intrinsic restriction for 
  the Rastall theory and perhaps  even for a large class of non-Lagrangian theories of gravity.

\section{Basic equations}

  The field equations of Rastall's theory where the only source of gravity is
  a scalar field $\phi$ with an arbitrary self-interaction potential $V(\phi)$
  can be written in the form\footnote
  		{The Rastall parameter $a$ is related to the other frequently used parameter 
  		$\lambda$ by $a = \dfrac{3\lambda-2}{2\lambda -1}$. }
\bearr                  \label{EE-G}
	R_{\mu\nu} - \frac{1}{2}g_{\mu\nu}R = \epsilon\biggr\{\phi_{;\mu}\phi_{;\nu} 
	- \frac{2 - a}{2}g_{\mu\nu}\phi^{;\rho}\phi_{;\rho}\biggl\} + (3 - 2a)g_{\mu\nu}V(\phi),
\yyy                    \label{e-phi}
	\Box\phi + (a - 1) \frac{\phi^{;\rho}\phi^{;\sigma}
	\phi_{;\rho;\sigma}}{\phi_{;\alpha}\phi^{;\alpha}} = - \epsilon(3 - 2a)V_\phi,
\ear
  where $a$ is the free parameter of Rastall's theory, and the equations of GR
  are restored in the case $a=1$; $\epsilon = \pm 1$, so that $\epsilon=+1$ 
  corresponds to a canonical scalar field, and $\epsilon = -1$ to a phantom one.
  
    Equations \rf{EE-G} can be rewritten as
\beq
	R_{\mu\nu} = \epsilon\biggr\{\phi_{;\mu}\phi_{;\nu} 
	+ \frac{1 - a}{2}g_{\mu\nu}\phi^{;\rho}\phi_{;\rho}\biggl\} - (3 - 2a)g_{\mu\nu}V(\phi),
\eeq

  We are going to discuss the linear stability of \ssph\ solutions of the theory under
  \sph\ perturbations.  Accordingly, we consider the general \sph\ metric
\beq                 \label {ds}
	ds^2 = e^{2\gamma} dt^2 - e^{2\alpha} du^2 - e^{2\beta} d\Omega^2,	
\eeq    
  where $\alpha, \beta, \gamma$ are functions of $u$ and $t$, and  
  $d\Omega^2$ is the line element on a unit sphere. We also assume 
  $\phi = \phi(u, t)$.
  
  Assuming that our perturbed system only slightly deflects from a static background
  solution, let us write the field equations, neglecting all terms nonlinear 
  with respect to time derivatives:
\bear
	- e^{2(\alpha - \gamma)} (\ddot\alpha + 2\ddot\beta) 
	+ \gamma''+ \gamma'(\gamma' - \alpha' + 2\beta') -
	 	&=& - \frac{\epsilon}{2}(1 - a)\phi'^2 - (3 - 2a)e^{2\alpha}V ,
\yy
        - e^{2(\alpha - \gamma)}\ddot\alpha  
         + \gamma'' + 2\beta'' - \alpha'(\gamma'+ 2\beta') + \gamma'^2 
                                          + 2\beta'^2  
         &=& - \frac{\epsilon}{2}(3 - a)\phi'^2 - (3 - 2a)e^{2\alpha}V ,
\yy
	- e^{2(\alpha - \gamma)}\ddot\beta  
	+ \beta'' + \beta'(\gamma' - \alpha'+ 2\beta') - e^{2(\alpha - \beta)} 
	&=& - \frac{\epsilon}{2}(1 - a)\phi'^2 - (3 - 2a)e^{2\alpha}V ,
\yy
       \dot\beta' + (\beta'- \gamma')\dot\beta - \beta'\dot\alpha  
		&=&  - \frac{\epsilon}{2}\phi'\dot\phi,
\yy
	a\phi'' + [\gamma'- a\alpha' + 2\beta']\phi'- e^{2(\alpha - \gamma)}\ddot\phi  
					&=&  \epsilon(3 - 2a)e^{2\alpha}V_\phi
\ear
  (the overdot stands for $\d/\d t$, the prime for $\d/\d u$, the subscript $\phi$ for $d/d\phi$).  
  For the static background we have the equations 
\bear
	\gamma''+ \gamma'(\gamma' - \alpha' + 2\beta') \eql
			- \frac{\epsilon}{2}(1 - a)\phi'^2 - e^{2\alpha}W, 
\yy
	\gamma'' + 2\beta'' - \alpha'(\gamma'+ 2\beta') + \gamma'^2 + 2\beta'^2 \eql
			- \frac{\epsilon}{2}(3 - a)\phi'^2 - e^{2\alpha}W, 
\yy
	\beta'' + \beta'(\gamma' - \alpha'+ 2\beta') - e^{2(\alpha - \beta)} 
		\eql - \frac{\epsilon}{2}(1 - a)\phi'^2 - e^{2\alpha}W,
\yy
	a\phi'' + [\gamma'- a\alpha' + 2\beta']\phi' \eql \epsilon  e^{2\alpha}W_\phi,
\ear
   where $W(\phi) = (3-2a) V(\phi)$.  Assuming, with some small parameter $\eps$, 
\[   
      \phi (u,t) = \phi(u) + \delta\phi (u,t),  \qquad  \delta\phi \sim \eps \ll 1,      
\]
  and similarly for all other variables, we can write the perturbation equations in
  the linear order $O(\eps)$:
\bearr
	- e^{2(\alpha - \gamma)}(\delta\ddot\alpha + 2\ddot\delta\beta)
	+  \delta\gamma''+ \delta\gamma'(\gamma' - \alpha' + 2\beta')
	+ \gamma'(\delta\gamma' - \delta\alpha' + 2\delta\beta') 
\nnn \inch \inch	
 	=  - \epsilon(1 - a)\phi'\delta\phi' - e^{2\alpha}(2\delta\alpha W + W_\phi\delta\phi),
\yyy
	- e^{2(\alpha - \gamma)}\delta\ddot\alpha 
	+ \delta\gamma'' + 2\delta\beta'' - \delta\alpha'(\gamma'+ 2\beta')
	- \alpha'(\delta\gamma'+ 2\delta\beta') + 2\delta\gamma\gamma' 
	+ 4\beta'\delta\beta' 
\nnn \inch \inch	
	=- \epsilon(3 - a)\phi'\delta\phi' - e^{2\alpha}(2\delta\alpha W + W_\phi\delta\phi),
\yyy
	- e^{2(\alpha - \gamma)}\delta\ddot\beta
	+\delta\beta'' + \delta\beta'(\gamma' - \alpha'+ 2\beta')
	 + \beta'(\delta\gamma' - \delta\alpha'+ 2\delta\beta')  
	 - 2e^{2(\alpha - \beta)} (\delta\alpha - \delta\beta)
\nnn \inch\inch
	= - \epsilon(1 - a)\phi'\delta\phi' - e^{2\alpha}(2\delta\alpha W + W_\phi\delta\phi),
\yyy \cm
	\delta\dot\beta' + (\beta'- \gamma')\delta\dot\beta - \beta'\delta\dot\alpha = - 		
			\frac{\epsilon}{2}\phi'\delta\dot\phi,
\yyy
	 - e^{2(\alpha - \gamma)}\delta\ddot\phi 
	 + a\delta\phi'' + [\gamma'- a\alpha' + 2\beta']\delta\phi' 
	 +  [\delta\gamma'- a\delta\alpha' + 2\delta\beta']\phi'
\nnn \inch\inch	 
	 = \epsilon  e^{2\alpha}(2\delta\alpha W_\phi + W_{\phi\phi}\delta\phi).
\ear

   These equations are written in the most genearl form and contain two kinds of
   arbitrariness: the choice of the radial coordinate $u$ in the background static
   metric and the perturbation gauge that fixes the reference frame in perturbed
   space-time. 

\section{Master equation and a discrepancy}

  As in GR, this system possesses only one dynamic degree of freedom connected 
  with the scalar perturbation $\delta\phi$. Accordingly, the perturbation equations 
  can be used to exclude the metric perturbations and to obtain a single 
  ``master equation'' for $\delta\phi$. This can be achieved most conveniently
  using the gauge $\delta\beta \equiv 0$. The perturbation equations become
\bearr	           \label{e00}
	 - e^{2(\alpha - \gamma)}\delta\ddot\alpha 
	+ \delta\gamma''+ \delta\gamma'(2\gamma' - \alpha' + 2\beta')- \gamma' \delta\alpha'
	=  - \epsilon(1 - a)\phi'\delta\phi' - e^{2\alpha}(2W \delta\alpha  + W_\phi\delta\phi),
\yyy		           \label{e11}
	- e^{2(\alpha - \gamma)}\delta\ddot\alpha
	+ \delta\gamma'' - \delta\alpha'(\gamma'+ 2\beta')- (\alpha' - 2\gamma')\delta\gamma' 
	=	- \epsilon(3 - a)\phi'\delta\phi' - e^{2\alpha}(2 W\delta\alpha + W_\phi\delta\phi),
\yyy	           \label{e22}
	\beta'(\delta\gamma' - \delta\alpha')  - 2e^{2(\alpha - \beta)} \delta\alpha 
	 = - \epsilon(1 - a)\phi'\delta\phi' - e^{2\alpha}(2W \delta\alpha  + W_\phi\delta\phi),
\yyy             \label{e01}	
	- \beta'\delta\dot\alpha = - \frac{\epsilon}{2}\phi'\delta\dot\phi,
\yyy	 	           \label{e-phi}
 	a\delta\phi'' + [\gamma'- a\alpha' + 2\beta']\delta\phi' + [\delta\gamma'- a\delta\alpha']\phi'
 	- e^{2(\alpha - \gamma)}\delta\ddot\phi 
		= \epsilon e^{2\alpha}(2W_\phi \delta\alpha  + W_{\phi\phi}\delta\phi).
\ear
  Equation \rf{e01} is easily integrated in $t$ leading to 	
\beq            \label{da}
	\delta\alpha = \eta(u) \delta\phi + \xi(u),
\eeq
  where $\xi(u)$ is an arbitrary function corresponding to possible static perturbations while 
  $\eta(u)$ is defined as	
\beq        \label{eta}
		\eta = \epsilon \phi'/(2\beta'). 
\eeq
   On the other hand, the difference of \rf{e00} and \rf{e11}) gives
\beq             \label{ga1}
	\delta\gamma' =\eta\delta\phi' - \eta'\delta\phi.
\eeq
  Ignoring static perturbations (that is, putting $\xi(u) \equiv 0$), a substitution 
  of \rf{da} and \rf {ga1} into \rf{e-phi} leads to the final master equation
\bearr                              \label{master}
	- e^{2(\alpha - \gamma)} \delta{\ddot\phi}
	+ a\delta\phi'' + \Big[ 2\beta' + \gamma' - a\alpha'+ \eta(1 - a)\phi'\Big]\delta\phi' 
\nnn	 \inch \inch
	+ \Big[- (1 + a)\eta'\phi' - \epsilon  e^{2\alpha}\big(2\eta W_\phi 
	+ W_{\phi\phi}\big)\Big]\delta\phi = 0,.
\ear
  whose analysis for particular solutions of the background equations should lead to definite
  conclusions on their stability or instability. 

  However, $\delta\gamma'$  may be alternatively calculated from \eqref{e22}, which gives
\beq               \label{ga2}
		2\beta' \delta\gamma' = \epsilon a \phi' \delta\phi' 
				+ e^{2\alpha}\Big[2 (e^{-2\beta}-W) \delta\alpha - W_\phi \delta\phi \Big].
\eeq    
  Comparing \eqref{ga1} and \eqref{ga2} and using the background equations, we obtain 
  the relation
\beq                            \label{gammas}
		(1-a) \phi' \delta\phi' = (1- a) \big[ \phi'' - \alpha' \phi' + \eta \phi'{}^2\big] \delta\phi
\eeq    
  (the calculation is most conveniently carried out using the harmonic radial coordinate, such that 
  $\alpha = 2\beta + \gamma$). We immediately see that at $a =1$, that is, in GR, this equation
  becomes an identity, thus confirming that the stability study in GR remains consistent. 
  However, with $a \ne 1$, in Rastall's theory, \eq \eqref{gammas} is nontrivial, and its 
  integration in $u$ gives
\beq                         \label{df}
			\delta\phi = F(t) \phi' e^{\alpha + Q(u)}, \qquad Q(u) = \int \eta(u) \phi'(u) du,
\eeq      
  where $F(t)$ is an arbitrary function of time, which only should be small, i.e., $O(\eps)$. 
  
  Now, we can substitute \eqref{df} to the master equation \eqref{master} and observe 
  the usual separation of variables: the quantity ${\ddot F}/F$  will be equal to a certain 
  combination of functions taking part in the background solution, hence this 
  combination is equal to some separation constant. Such a condition, is, in general, not 
  satisfied by the background solution, which makes the whole stability study inconsistent.  
  
  We will confirm this observation using two known background solutions of Rastall theory 
  as examples. Using one of them, we will demonstrate that invoking a nonzero $\xi(u)$ in 
  \rf{da} does not solve the consistency problem. 

\section{Special solutions}
\subsection{$a = -1$, $V = 0$}

  In this case, the solution has been obtained using the quasiglobal radial coordinate 
  such that $\alpha = - \gamma$, and the background equation are 
\bear
	\gamma''+ 2\gamma'(\gamma'+ \beta') &=&  - \epsilon\phi'^2 ,
\yy
	\gamma'' + 2\beta'' + 2\gamma'(\gamma'+ \beta') +  2\beta'^2 &=& - 2\epsilon\phi'^2 ,
\yy
	\beta'' + 2\beta'(\gamma' + \beta') - e^{-2(\gamma + \beta)} &=& - \epsilon\phi'^2,
\yy	
	- \phi'' + 2\beta'\phi' &=& 0.
\ear
  The background solutions are possible only for $\epsilon = -1$ and are given by 
  (denoting now the radial coordinates by $x$)
\bearr
	ds^2 = A(x)dt^2 - \frac {dx^2}{A(x)} - \sqrt{\frac{3}{2C^2}}\frac{d\Omega^2}{x},
\yyy
	A(x) = {K}/{x} -  (C/\sqrt{6})\,x^3,
\yyy
		\phi =  \sqrt{3/2} \ln x + \phi_0,
\ear
  where $\phi_0$, $C$ and $K$ are integration constants. The metric has the same properties 
  as in the k-essence case studied in Ref. \cite{kirillk}, with a horizon at 
  $x = \biggr(\dfrac{\sqrt{6}K}{C}\biggl)^{1/4}$ and a singular spatial infinity $x \to \infty$. 
  Unlike \cite{kirillk}, however, the scalar field is here defined as to span from minus to plus infinity.

   The master equation \rf{master} now reduces to
\beq             \label{mas-1}
	  e^{-4\gamma} \delta{\ddot \phi}  + \delta\phi''  + \frac 2x \delta\phi'  =0.
\eeq
  Two different expressions for $\delta \gamma'$, obtained as described above, are
\beq
	     \delta \gamma' = \sqrt{3/2} \delta \phi' = - \sqrt{3/2} \delta \phi' 
	     				- \sqrt{6} e^{2\alpha - 2\beta} \delta \phi.	
\eeq    
  Substituting the background solution, we are able to integrate the last equality in $x$,
  obtaining
\beq                                                   \label{df--1}
		\delta \phi = F(t) \psi(x), \qquad \psi(x) = (K - (C/\sqrt{6})x^4)^{-1/2}.
\eeq    
   Substituting \eqref{df--1} into \eqref{mas-1} and separating the variables, we obtain
\beq
			- \frac{\ddot F}{F} 	= \frac{\psi''}{\psi} + \frac 2x \frac{\psi'}{\psi} = \const.	 		 		
\eeq        
   It is easy to verify that the last equation does not hold for $\psi(x)$ given by \eqref {df--1}. 
  
\subsection{$a = 0$, $V = \Lambda = \const$}

  We are using here the  the harmonic radial coordinate, such that
   $\alpha =2\beta + \gamma$.  The background equations are
\bear
	\gamma'' &=& - \frac{\epsilon}{2}\phi'^2 - 3e^{2\alpha}\Lambda,
\yy
	\gamma'' + 2\beta'' - \alpha'^2 + \gamma'^2 + 2\beta'^2 
		&=& - \frac{3}{2}\epsilon\phi'^2 - 3e^{2\alpha}\Lambda,
\yy
	\beta'' - e^{2(\gamma + \beta)} &=& - \frac{\epsilon}{2}\phi'^2 - 3e^{2\alpha}\Lambda,
\yy
	\alpha'\phi' &=& 0.
\ear
  The last equation implies $\alpha =\const$, which can be rescaled to $\alpha = 0$ by a trivial 
  redefinition of the radial coordinate.  The solution has the form
\bearr
	ds^2 = \frac{9b^4}{\cosh^4bu}dt^2 - du^2 - \frac{\cosh^2bu}{3b^2}d\Omega^2,
\yyy
	\phi' = \pm b\sqrt{6 - \frac{4}{\cosh^2 bu}}, \qquad \epsilon = - 1, \qquad
	b = \sqrt{\Lambda}.
\ear

  Equation \rf{e-phi} for scalar field perturbations now reads
\beq                        \label{e-phi0} 
	        e^{- 2\gamma}\delta\ddot\phi -\phi'\delta\gamma' = 0.
\eeq    

  Two alternative expressions for $\delta\gamma'$ are \rf{ga1} and, as given by \rf{ga2} after
  using the background solution,
\beq                \label{ga3}
		\delta\gamma' = \frac{3}{2}\phi'\delta\phi.
\eeq
  Comparing \rf{ga1} and \rf{ga3}, we obtain 
\beq
		\eta \delta\phi'- \eta' \delta\phi = \frac{3}{2}\phi'\delta\phi.
\eeq
  We can integrate this equation, finding the spatial behavior of $\delta\phi$:
\beq
		\delta\phi = F(t) \frac{\phi'}{\beta'}e^{-3\beta}, \qquad F(t) = \mbox{arbitrary function.}
\eeq
  Substituting it to the master equation \rf{e-phi0} with $\delta\gamma'$  given by \rf{ga3}, we
  arrive at
\beq
		\frac 23 \frac{\ddot F}{F} = \phi'{}^2 e^{-4\beta} = \const.
\eeq        
   The latter equality does not hold for our background solution. 
   
   We can ask if we can cure this problem by considering the ``integration constant'' $\xi(u)$
   that appears in \rf{da}. Then the two expressions for $\delta\gamma'$ become
\beq
     \delta\gamma' = \eta \delta\phi'- \eta' \delta\phi - \xi' = \frac{3}{2}\phi'\delta\phi - 3\beta' \xi.
\eeq
   This relation is integrated giving 
\beq
	  \delta \phi = \eta  e^{-3\beta} [F(t) + H(u)], \qquad 
	                 	H(u) = \int \frac{e^{3\beta}}{\eta^2}(\xi' - 3\beta'\xi) du,
\eeq  
  where $H(u)$ is actually an arbitrary function of $u$ due to arbitrariness of $\xi$.
  Inserting this $\delta \phi$ to the master equation, we obtain\
\beq                           \label{xixi} 
	  \ddot F - \frac{3}{2}\phi'^2 e^{-4\beta} [F (t) + H(u)] - 6\beta'^2e^{-\beta}\xi = 0.	   
\eeq
  Differentiating \rf{xixi} with respect to $u$, we obtain a relation between functions of $u$ 
  (making it possible to calculate $\xi(u)$ in terms of the background functions) and also 
  that $F = \const$. As could be expected, the assumption $\xi(u) \ne 0$ leads to
  a purely static perturbation of the background, and time-dependent perturbations turn 
  out to be impossible. Thus invoking $\xi(u)$ only produces a static perturbation  but 
  does not solve the consistency problem for time-dependent perturbations.
  
\section{Conclusion}  

  We have performed a stability analysis of the exact black hole-type solutions found in the 
  context of the Rastall theory of gravity in the presence of a self-interacting scalar field,
  which were originally reported in Ref. \cite{kirillr}. The corresponding metrics are the same 
  as those  found in the context of k-essence theories \cite{kirillk}. This coincidence of the
  metrics found in so different contexts (but with quite different behaviors of the corresponding
  scalar fields)  was discussed in detail in Ref. \cite{kirille}. The perturbation analysis of the 
  k-essence solutions was carried out in \cite{kirills}, and it was concluded that they are 
  unstable. Hence, it was natural to perform a similar analysis for the corresponding Rastall
  solutions since the equivalence with the k-essence solutions may not be preserved at 
  the perturbative level.

  A surprising results is that such a stability analysis in the Rastall case is inconsistent: 
  linear time-dependent spherically symmetric perturbations simply do not exist. 
  In this sense, the solutions may be said to be stable under such perturbations.
  The reason for our conclusion is that in the Rastall theory different combinations of the 
  perturbed equations lead to different ``master'' equations for  the scalar field perturbation
  $\delta\phi$. This problem does not exist for k-essence solutions. We cannot attribute 
  this feature to a wrong choice of the coordinates or the perturbation gauge since, as 
  discussed, e.g., in \cite{mex,kz,kb18}, the perturbation method employed here is equivalent 
   to a gauge-invariant method.

  Very probably the roots of the inconsistency detected for Rastall black hole-type solutions
  come from the absence of a Lagrangian formulation of this theory. It must be remarked that 
  a similar inconsistency was found in the cosmological context, also with a scalar field as 
  a matter source. But, for the cosmological solutions  the inconsistency has been cured 
  by introducing ordinary baryonic matter. In the static, spherically symmetric case studied in 
  \cite{kirillr} such an extension is less obvious. 
  
  There have been some attempts to find a Lagrangian formulation for the Rastall theory, see, 
  e.g.,. \cite{smalley,jose}, but the resulting theories were not completely equivalent to the 
  Rastall one, or requires a completely new geometric framework: using these formulations,
  we, strictly speaking, go away from the original context of the Rastall theory.

  The results reported here may point out at some restrictions inherent to the applicability
  of the Rastall theory, and even maybe of any non-Lagrangian theory. We hope to extend 
  the present analysis in order to try to answer this question in our future works.

\subsection* {Acknowledgments:} 

  J.C. Fabris and O.F. Piattella thank CNPq (Brazil) and FAPES (Brazil)
  for partial financial support. D.C. Rodrigues was financed in part by the Coordena\c{c}\~ao de
  Aperfeitoamento de Pessoal de N\'{\i}­vel Superior - Brasil (CAPES) - Finance Code 001. 
  E.C.O. Santos thanks FAPES (Brazil) for financial support.
  The work of K. Bronnikov was supported by the RUDN University program 5-100 and by the 
  Russian Foundation for Basic Research Project 19-02-00346. 
  The work of K.B. was also performed within the framework of the Center FRPP 
  supported by MEPhI Academic Excellence Project (contract No. 02.a03.21.0005,
  27.08.2013).

\end{document}